# Far-field optical microscope with nanometer-scale resolution based on in-plane surface plasmon imaging.

Igor I. Smolyaninov

*Department of Electrical and Computer Engineering, University of Maryland, College Park, MD 20742, USA*

**ABSTRACT**

**A new far-field optical microscopy technique capable of reaching nanometer-scale resolution has been developed recently using the in-plane image magnification by surface plasmon polaritons. This microscopy is based on the optical properties of a metal-dielectric interface that may, in principle, provide extremely large values of the effective refractive index $n_{eff}$ up to $10^2$-$10^3$ as seen by the surface plasmons. Thus, the theoretical diffraction limit on resolution becomes $\lambda /2n_{eff}$, and falls into the nanometer-scale range. The experimental realization of the microscope has demonstrated the optical resolution better than 50 nm for 502 nm illumination wavelength. However, the theory of such surface plasmon-based far-field microscope presented so far gives an oversimplified picture of its operation. For example, the imaginary part of the metal's dielectric constant severely limits the surface-plasmon propagation and the shortest attainable wavelength in most cases, which in turn limits the microscope magnification. Here I describe how this limitation has been overcome in the experiment, and analyze the practical limits on the surface plasmon microscope resolution. In addition, I present more experimental results, which strongly support the conclusion of extremely high spatial resolution of the surface plasmon microscope.**

## I. INTRODUCTION.

Far-field optical microscopy remains invaluable in many fields of science, even though various electron and scanning probe microscopes have long surpassed it in resolving power. The main advantages of the far-field optical microscope are the ease of operation and direct sample visualization. Unfortunately, the resolution of a regular optical microscope is limited by the wavelength of visible light. The reason for the limited resolution is diffraction and, ultimately, the uncertainty principle: a wave can not be localized much tighter than half of its vacuum wavelength $\lambda/2$. Immersion microscopes [1] introduced by Abbe in the 19th century have slightly improved resolution on the order of $\lambda/2n$ because of the shorter wavelength of light $\lambda/n$ in a medium with refractive index $n$. However, immersion microscopes are limited by the small range of refractive indices $n$ of available transparent materials. For a while it was believed that the only way to achieve nanometer-scale spatial resolution in an optical microscope is to beat diffraction, and detect evanescent optical waves in very close proximity to a studied sample using a scanning near-field optical microscope [2]. Although many fascinating results are being obtained with near-field optics, such microscopes are not as versatile and convenient to use as regular far-field optical microscopes. For example, an image of a near-field optical microscope is obtained by point-by-point scanning, which is an indirect and a rather slow process.

Very recently it was realized [3] that a dielectric droplet on a metal surface which supports propagation of surface plasmons (or surface plasmon polaritons) may have an extremely large effective refractive index as seen by these modes. The properties of surface plasmons and convenient ways to excite them are described in detail in [4]. The wave vector of a surface plasmon propagating over an interface between a dielectric and an infinitely thick metal film is defined by the expression

$$k_p = \frac{\omega}{c}\left(\frac{\varepsilon_d \varepsilon_m}{\varepsilon_d + \varepsilon_m}\right)^{1/2} \qquad (1)$$

where $\varepsilon_m(\omega)$ and $\varepsilon_d(\omega)$ are the frequency-dependent dielectric constants of the metal and dielectric, respectively. If the imaginary part of the metal's dielectric constant is neglected, under the resonant condition

$$\varepsilon_m(\omega) = -\varepsilon_d(\omega) \qquad (2)$$

both phase and group velocities of the surface plasmons tend to zero. This means that the wavelength $\lambda_p$ of such modes becomes very small just below the optical frequency defined by equation (2), or in other words, the effective refractive index of the dielectric $n_{eff}$ becomes extremely large as seen by the propagating surface plasmons in this frequency range. As a result, a small droplet of liquid dielectric on the metal surface becomes a very strong lens for surface plasmons propagating through the droplet from the outside. On the other hand, the droplet boundary becomes an extremely efficient mirror for surface plasmons propagating inside the droplet at almost any angle of incidence due to the total internal reflection (this leads to the "black hole" analogy described in [3]).

This realization has led to the introduction of a surface plasmon immersion microscope [5], which has been recently described in [6]. Let us consider a far-field two-dimensional (2D) optical microscope made of such droplets or any other appropriately shaped planar dielectric lenses and/or mirrors as shown in Fig.1(a). Since the wavelength of surface plasmons $\lambda_p$ observed in the STM light emission [7] and near-field optical experiments [8] may be as small as a few nanometers (hence $n_{eff} = \lambda/\lambda_p$ may reach extremely large values up to $10^2$), the diffraction limit of resolution of such a 2D microscope may approach $\lambda_p/2$ or $\lambda/2n_{eff}$ . Theoretically it may reach a scale of a few nanometers. If a sample under investigation is forced to emit propagating surface plasmons, or if it is illuminated by propagating plasmons, these plasmons may produce a 2D magnified image of the sample in the appropriate location on the metal surface. Because of the metal surface roughness and the Raleigh scattering in the dielectric, the propagating plasmons are constantly scattered into normal photons propagating in free space. As a result, the plasmon-produced far-field 2D image on the metal surface may be visualized by a normal optical microscope. The image brightness far exceeds the background of scattered plasmons in other areas of the 2D microscope, and in addition, a fluorescence scheme of surface plasmon field visualization by a far-field optical microscope may be used [9]. The exact coupling efficiency between the plasmon-produced image and photons in free space which may be collected by a regular microscope depends on the surface roughness and/or the type of fluorescent dye used in the microscope. A typical surface plasmon resonance linewidth measured in the experiment is in the 1-10 % range [4], which indicates plasmon to photon conversion efficiency due to surface roughness of about the same order of magnitude. About the same conversion efficiency has been observed in the fluorescent imaging experiment [9]. In addition, this coupling efficiency may be improved by introducing an artificial periodic corrugation of the metal surface (however, such an artificial surface corrugation may cause difficulties in distinguishing real objects from the patterns produced by periodic corrugation).

Thus, the goal of a 2D microscope design is to have sufficiently high 2D image magnification, so that all the 2D image details would be larger than the $\lambda/2$ resolution limit of the normal optical microscope. As a result, a far-field optical microscope with nanometer-scale resolution would be produced. Very recently a microscope built on this principle has been reported [6], and experimental proofs of its resolution of at least 50 nm, which is equal to approximately $\lambda/10$ and far supersedes resolution of any other known far-field optical microscope have been presented. We believe that this new microscopy technique will lead to numerous breakthroughs in biological imaging and subwavelength lithography. However, the theoretical description of the microscope given above presents an oversimplified picture of the microscope operation. For example, the imaginary part of the metal's dielectric constant severely limits the shortest attainable surface plasmon wavelength and the surface-plasmon propagation length in most cases. This in turn limits the microscope's 2D magnification in the metal plane. In this paper I will describe how these limitations have been overcome in the experiment, and analyze the practical limits on the surface plasmon microscope resolution. In addition, I will present more experimental results, which strongly support the conclusion of extremely high spatial resolution of the surface plasmon microscope.

## II. HOW SHORT MAY BE THE SURFACE PLASMON WAVELENGTH?

The amplitude of every resonance in nature is limited by the energy losses. The same statement is valid with respect to the surface plasmon resonance. It is clear from eq.(1) that the imaginary part of $\varepsilon_m(\omega)$ limits the shortest attainable wavelength of surface plasmons on an infinitely thick metal film. If we assume that $\varepsilon_d$ is real, while $\varepsilon_m=\varepsilon^{(r)}_m+i\varepsilon^{(i)}_m$, the shortest wavelength of a surface plasmon would be equal to

$$\lambda_{P\min} = \lambda\left(-\frac{2\varepsilon_m^{(i)}}{\varepsilon_d\varepsilon_m^{(r)}}\right)^{1/2} \qquad (3)$$

In the frequency range of the Ar-ion laser lines (which corresponds to the plasmon resonance at the gold-glycerin interface used in [6]) this value could not be much smaller then 200 nm. Thus, the idealized surface plasmon dispersion curve shown in Fig.1(b) has nothing to do with reality if the gold film is very thick and glycerin is used as a dielectric. However, the situation changes radically if the gold film thickness falls into the few tens of nanometers range, and the dielectric constant of the substrate used for the gold film is chosen to coincide with the dielectric constant of the liquid droplet on the gold film surface. In such a case a pair of surface plasmon modes appears [10] (the symmetric and the antisymmetric solutions of the Maxwell equations), and in the large wave vector limit the surface plasmon dispersion in eq.(1) is modified to look as follows:

$$k_p = \frac{\omega}{c}\left(\frac{\varepsilon_d \varepsilon_m}{\varepsilon_d + \varepsilon_m \pm 2\varepsilon_d e^{-k_p d}}\right)^{1/2} , \qquad (4)$$

where $d$ is the gold film thickness. The term $2\varepsilon_d e^{-k_p d}$ in the denominator of eq.(4) has real and imaginary parts, so that by playing with the frequency and the gold film thickness the plasmon momentum may be forced to diverge in the case of the antisymmetric plasmon mode [10]. As a result, the use of an idealized surface plasmon dispersion curve shown in Fig.1(b) is justified for a finite thickness of the gold film in a situation in which the dielectric constants of the droplet and the substrate are close to each other. Thus, glycerin with the refractive index of $n_{gl}$ = 1.47 is ideally suited for experiments performed with gold films deposited onto a glass substrate. I should also mention that the use of materials with larger dielectric constants in the visible range (such as diamond or semiconductors) would improve the situation even for the thicker gold films (and make it more close to an ideal) since this would shift the plasmon resonance towards longer wavelengths where the $\varepsilon^{(i)}{}_m$ falls very rapidly.

## III. HOW TO EXTEND THE SURFACE PLASMON PROPAGATION LENGTH?

While the use of idealized surface plasmon dispersion curve in Fig.1(b) for glycerin droplets has been justified in the previous section, even more crucial question for the consideration of the far-field surface plasmon microscope performance is the surface plasmon propagation length. The importance of this question may again be illustrated in the case of infinitely thick metal film. For a complex $\varepsilon_m$ the imaginary part of $k_p$ from eq.(1) determines the surface plasmon propagation length $L_p$. Around $\lambda_{Pmin}$ the propagation length becomes extremely short: $L_p \sim 2\lambda_{Pmin}$ , and it is clear that a far-field surface plasmon microscope could not be built in this case. However, it appears that the use of symmetric geometry may again help to overcome the surface plasmon propagation problem. The effect of dramatic enhancement of the surface plasmon propagation length over a thin metal film in the symmetric configuration has been described previously by Burke *et al.* in [10]. According to their calculations, the plasmon propagation over a symmetric structure appears to be typically an order of magnitude larger compared to the case of an asymmetric structure. For example, a surface plasmon propagation at $\lambda$=633 nm over a 15 nm thick silver film surrounded on both sides by a dielectric with refractive index 1.5 may reach 610 micrometers. Moreover, Burke *et al.* had found two additional leaky surface-plasmon-like solutions in the thin film geometry and noted that such leaky modes may even grow in intensity with distance under the resonant excitation if the rate of energy influx from the excitation source is greater than the dissipation in metal.

Here we should point out that in our experiments [6] a substantial portion of surface plasmon propagation occurs over the areas of gold films which were perforated by the periodic arrays of nanoholes. It is clear that all the surface plasmon-like modes, which propagate over a periodically corrugated gold surface must be leaky modes due to the photonic crystal effects. The dispersion laws of surface plasmons and normal photons propagating inside the dielectric at small angles along the metal-dielectric interface are shown in Fig.2, where *k* represents the quasi-momentum of the respective electromagnetic mode. Since each branch (photon or plasmon) of the dispersion law can be shifted along the k-axis by an integer number of the inverse lattice vectors, it is clear that these branches have an infinite number of intersections with each other. These intersections are shown by the dots in Fig.2. Irrespective of the nature of the periodic corrugation (nanoholes like in our experiments, or something else), the propagation length of surface-plasmon-like modes drastically changes near these intersection points. According to the observation by Burke *et al.* in [10] plasmon propagation length near the intersection points between the dispersion laws of plasmon-like modes and photons in the dielectric should increase dramatically. The physical reason for this effect may be understood as though plasmons spend some of their lifetime as regular photons, and thus, propagate much farther. On the other hand, under the resonant excitation plasmon-like leaky modes which propagate over a periodic surface may even grow in intensity if the rate of energy influx from the excitation source is greater than the dissipation in metal. I should also point out that the vast majority of the intersection points in Fig.2 are located in the large wave vectors area of the unperturbed plasmon dispersion curve, which is exactly the property needed for high resolution microscopy. Thus, while the exact values of surface plasmon propagation length over a periodic nanohole array need to be calculated from the first principles, there exist good reasons for this propagation length to be large over the nanohole array for short-wavelength plasmons. In order to achieve the best possible magnification of the plasmon microscope both effects of the plasmon propagation length increase described above should be used to our advantage: the preferred geometry of the 2D microscope should be based on a thin periodically corrugated metal film surrounded on both sides by dielectric media with equal dielectric constants. The results of our measurements of surface plasmon propagation length shown in Fig.3 and described in detail in Section V confirm substantial increase of the surface plasmon propagation length in a symmetric configuration chosen in our experiments.

## IV. THE ROLE OF MODE COUPLING.

Liquid droplets with large-enough thickness may support not only the surface plasmons at the metal-dielectric interface but regular guided modes as well (Fig.1(b)). These guided modes are similar to the electromagnetic modes that propagate in dielectric waveguides. The droplet profile changes with distance along its optical axis: ideally the droplet has a parabolic shape in the *xy*-plane, and in addition, the droplet thickness varies in the *z*-direction. Far from the droplet edges both the droplet thickness and the droplet width vary slowly, which leads to weak coupling between all the electromagnetic modes of the system due to momentum non-conservation (because of the loss of translation symmetry along the metal plane). This effect has been observed in our experiments (see Fig.3(c,d) and detailed discussion in Section V). It may be extremely beneficial for the plasmon microscope performance.

The diffraction-limited angular resolution ~$\lambda_p/F$ of the microscope is defined by the plasmon propagation around the focal point of the parabolic mirror/droplet, where $F$ is the focal length of the mirror and $\lambda_p$ is the plasmon wavelength. Once the short-wavelength surface plasmons left the area in the vicinity of the focal point, and reached some more distant area of the droplet with a larger width $D>>F$, plasmon conversion into the guided modes with larger wavelength $\lambda_g$ may not lead to the deterioration of the angular resolution (see the sketch in Fig.3(e)). If $\lambda_p/F\sim\lambda_g/D$ angular resolution of the microscope will be conserved. After such a conversion, the two-dimensional image formed by the propagation of the guided modes will keep all the spatial information which would be contained in a plasmon-formed image if the plasmons would reach the geometrical location of the image. This statement is true as long as the geometrical optics description of the mode propagation inside the droplet remains valid, or if $\lambda_g<<F$. Thus, the mode coupling effect provides another way of solving the problem of short plasmon propagation length, which has been discussed in the previous section. Based on the discussion above, the best shape of the dielectric droplet seems to be a compound shape, which may be approximated by two parabolas such that the focal length of the first parabola is much smaller than the focal length of the second one. In this case the role of the parameter $D$ is played by the focal length of the second parabola, and the short-wavelength plasmons need to travel only a distance of the order of the focal length $F$ of the first one. Such a compound droplet shape has been used in some of our experiments described below.

## V. NEW EXPERIMENAL EVIDENCE OF SUPERRESOLUTION.

In a scheme similar to one described earlier in [3], glycerin microdroplets have been used as 2D optical elements in the design of the plasmon microscope. The dielectric constant of glycerin $\varepsilon_g$=2.161 is ideally suited for experiments performed on a gold surface within the wavelength range of the laser lines of an argon-ion laser (Fig.1(b)). At the $\lambda_0$=502 nm line the real part of the gold dielectric constant is $\varepsilon_m$= - 2.256 [11]. According to equation (1) the corresponding surface plasmon wavelength inside glycerin is $\lambda_p \sim 70$ nm, and the effective refractive index of glycerin is $n_{eff} = \lambda_0/\lambda_p \sim 7$. On the other hand, the use of glycerin achieves good dielectric constant matching with the silica glass, which has been used as a substrate for the gold films. According to the discussion in the previous chapters, this fact is important for improving surface plasmon propagation over the gold films with the thickness in the 50-100 nm range used in our experiments. The plasmon propagation length over the gold-glycerin interface at 502 nm has been measured using two complementary techniques: near-field imaging technique described in [3,12] and the fluorescent surface plasmon imaging technique similar to the one described in [9]. Both techniques gave similar results. In our experiments artificial pinholes in gold film were produced inside a thin glycerin droplet (which was stained with the bodipy dye) by touching the gold film with a sharp STM tip. Such pinholes are known to emit propagating surface plasmon beams [12]. The characteristic exponentially decaying surface plasmon beam (excited from the right side of the image) observed in this experiment is shown in Fig.3(a), which has been obtained using fluorescent imaging. The cross section of this beam shown in Fig.3(b) has been fitted by an exponent and indicate plasmon propagation length of the order of 3 micrometers at 502 nm laser wavelength. In some cases we were also able to image the process of surface plasmon coupling into the regular guided modes described in Section IV, as shown in Figs.3(c,d). Image (c) and its cross-section (d) show the effect of mode coupling due to the slowly varying shape of the glycerin droplet: quickly decaying surface plasmon beams emitted by two pinholes give rise to weaker guided mode beams, which exhibit much slower decay and longer propagation length.

In our microscopy experiments the samples were immersed inside glycerin droplets on the gold film surface. The droplets were formed in desired locations by bringing a small probe Fig.4(a) wetted in glycerin into close proximity to a sample. The probe was prepared from a tapered optical fiber, which has an epoxy microdroplet near its apex. Bringing the probe to a surface region covered with glycerin led to a glycerin microdroplet formation under the probe (Fig.4b). The size of the glycerin droplet was determined by the size of the seed droplet of epoxy. The glycerin droplet under the probe can be moved to a desired location under the visual control, using a regular microscope. Our droplet deposition procedure allowed us to form droplet shapes, which were reasonably close to parabolic. In addition, the liquid droplet boundary may be expected to be rather smooth because of the surface tension, which is essential for the proper performance of the droplet boundary as a 2D plasmon mirror. Thus, the droplet boundary was used as an efficient 2D parabolic mirror for propagating surface plasmons excited inside the droplet by external laser illumination.  Since the plasmon wavelength is much smaller than the droplet sizes, the image formation in such a mirror can be analyzed by simple geometrical optics in two dimensions.

Periodic nanohole arrays first studied by Ebbesen *et al.* [13] appear to be ideal test samples for the plasmon microscope. Illuminated by laser light such arrays produce propagating surface waves, which explains the anomalous transmission of such arrays at optical frequencies. Fig.5 shows various degrees of 2D image magnification obtained with a 30x30 μm$^2$ rectangular nanohole array with 500 nm hole spacing described in [14] and used as a test sample. In general, smaller glycerine droplets produced higher magnification in the images. It should be pointed out that all the guided modes in the droplet (surface plasmons and the regular guided modes shown in Fig.1(b)) participate in the formation of the 2D images. The relative contribution to the image of each mode changes with distance from the imaged sample due to varying mode coupling and decay. Approximate reconstructions of the images using 2D geometrical optics (via ray tracing) are shown next to each experimental image. If the shape of the 2D mirror (the droplet edge) is given by the exact parabolic dependence as $Y=X^2/2P$, the point $(X_1,Y_1)$ is reflected into the point $(X_2,Y_2)$ according to the following expressions:

$$x_2 = -\frac{P}{x_1}\left(\sqrt{\left(y_1 - \frac{P}{2}\right)^2 + x_1^2} - \left(y_1 - \frac{P}{2}\right)\right) \quad (5)$$

$$y_2 = \left(\frac{P^2}{2x_1^2} - \frac{1}{2}\right)\left(\sqrt{\left(y_1 - \frac{P}{2}\right)^2 + x_1^2} - \left(y_1 - \frac{P}{2}\right)\right) + \frac{P}{2} \quad (6)$$

These expressions are precise. However, the droplet shapes in our experiments may only approximately be represented by parabolas, and the damping of surface plasmon field over varying propagation lengths has not been included in the simulation (extensive sets of data on the plasmon propagation length versus the plasmon frequency and the metal and dielectric film thicknesses can be found in [10]). These facts limit the precision of our image reconstructions. Nevertheless, we have achieved an impressive qualitative agreement between the experimental and theoretical images of the plasmon microscope. In all the calculated images described below the individual nanoholes of the arrays are shown as individual dots in the theoretical images. Comparison of Fig.5(c) and Fig.5(d) indicates that the rows of nanoholes separated by 0.5 μm may have been resolved in the image (c) obtained using only a 10x objective of the conventional microscope, while comparison of Fig.5(e) and Fig.5(f) obtained using a 50x objective indicates that individual 150 nm diameter nanoholes separated by 0.5 μm gaps are resolved in the image (e) obtained at 502 nm. These individual nanoholes are located in close proximity to the focus of the droplet/mirror, and hence experience the highest image magnification. In fact, the image in Fig.5(e) shows successful use of the droplet with compound parabolic geometry described in the end of the previous section, which is supposed to take the full advantage of the mode coupling mechanism described in section IV. Even though the exact role of mode coupling in formation of each image in Fig.5 is not clear, it seems certain that the 2D images in Figs.5(a,c) are formed with considerable participation of the guided modes, since the distance travelled by the electromagnetic modes is of the order of 100 micrometers in these cases. While the image in Fig.5(a) does not contain any evidence of high resolution, the image in Fig.5(c) seems to demonstrate that the mode coupling does preserve high angular and spatial resolution, as has been discussed in section IV.

Another resolution test of the microscope has been performed using a 30x30 $\mu m^2$ array of triplet nanoholes (100 nm hole diameter with 40 nm distance between the hole edges) shown in Fig.6(c). This array was imaged using a glycerine droplet shown in Fig.6(a). The image of the triplet array obtained at 515 nm using a 100x microscope objective is shown in Fig.6(b) (compare it with an image in (d) calculated using the 2D geometrical optics). Although the expected resolution of the microscope at 515 nm is somewhat lower than at 502 nm, the 515 nm laser line is brighter, which allowed us to obtain more contrast in the 2D image. The least-distorted part of the image 6(b) (far from the droplet edge, yet close enough to the nanohole array, so that surface plasmon decay does not affect resolution) is shown at higher digital zooms of the CCD camera mounted onto our conventional optical microscope in Figs.6(e,f). These images clearly visualize the triplet nanohole structure of the sample.

According to the geometrical optics picture of the 2D plasmon microscope operation, its magnification $M$ is supposed to grow linearly with distance along the optical axis of the droplet/mirror:

$$M = \frac{2y}{P} - 1, \qquad (7)$$

where $P$ is the focal distance of the parabola. Our measurements of the image magnification indeed exhibit such linear dependence (Fig.7a). The dots in the graph show the distance between the neighbouring triplets in the image as a function of triplet position measured along the optical axis of the droplet. At small distances individual nanoholes are not resolved within the triplet. At larger distances (where the triplets are resolved, see the cross section in Fig.7(b) measured through the line of double holes in the image of the triplet array) the data points represent the positions of the triplet's centres. The gap in the data corresponds to the intermediate area of the image in which the feature identification in the image is difficult. The slope of the measured linear dependence in Fig.7(a) corresponds to P=7 μm , which is in reasonable agreement with the value of P of the order of 10 μm , which can be determined from the visible droplet dimensions in Fig.6(a).

While the simple geometrical optics model of the image formation agrees reasonably well with the experiment, a few alternative mechanisms may form an image of a periodic source, such as the Talbot effect [15]. However, resolution of the Talbot images also approximately equals to $\lambda/2n$. Thus, whatever optical mechanism is involved in the formation of the images of the triplets in Fig.6, short-wavelength plasmons are necessarily involved in this mechanism. In addition, reconstruction of the source image in the Talbot effect happens at the specific planes where exact field distribution of the source is reproduced. These planes are called the Talbot planes. At all the distances other then the set of Talbot distances the pattern of illumination differs greatly from the pattern of the source: instead of triad features of the source one may see sets of 6, 9, 12, etc. bright illumination maxima. This diffraction behavior is further complicated by the fact that different triads of the source are located at different distances from a given triad of the image. Since it is very hard to imagine that the periodicity of the source would exactly coincide with the periodicity of the Talbot planes spacing, the mechanism of image formation due to diffraction effects seems highly improbable. At the same time, all the diffraction and interference phenomena reproduce the geometrical optics description in the limit of small wavelengths. This fact is reflected in rather good agreement between the experimental images and the images calculated in the geometrical optics approximation.

In order to prove that the plasmon microscope is capable of aperiodic samples visualization, we have obtained images of small gaps in the periodic nanohole arrays (Fig.8). The electron microscope image of one of the gaps in the periodic array of nanoholes is shown in Fig.8(a). Two wider mutually orthogonal gaps were made in the array along both axis of the structure [14] as shown in the theoretical reconstruction in Fig.8(c). The plasmon image in Fig.8(b) and its cross section in Fig.8(d) obtained at 502 nm wavelength shows both the periodic nanohole structure and the gap in the structure indicated by the arrows in the images. The width of the gap in the image grows linearly with the distance from the sample in agreement with our measurements in Fig.7. In principle, the observed gap in the image might be interpreted as a Moire pattern, due to two shifted diffraction patterns from the two portions of the nanohole array separated by the gap. However, the cross section through the gap in the image (Fig.8d) may be considered as evidence against such interpretation. Dark stripes in the Moire patterns normally exhibit slightly attenuated brightness compared to the original overlapping illumination patterns. The contrast in the image between the gap and the images of nanoholes seems to be too large for a Moire pattern interpretation.

Finally, in order to evaluate the microscope resolution at the optimized 502 nm wavelength we have analyzed the cross-sections of the images of the triplet structure (similar to the one described earlier in Fig.6) obtained at this wavelength. The most magnified triplets, which are still discernible in the experimental image in Fig.9(a) are shown by the arrow (compare this image with the theoretical one shown in Fig.9(b)). These triplets are shown at a higher zoom in Fig.9(c). The cross section through two individual nanoholes in the triplet clearly shows the 40 nm gap between the nanoholes. While optical properties of this particular triplet may slightly differ from the designed values and lead to an appearance of a wider gap in the image, the distance between the centres of the nanoholes should be 140 nm. The cross section in Fig.9(c) seems to indicate at least three times better resolution of the plasmon microscope of about 50 nm. Thus, at least 50 nm ($\lambda/10$) spatial resolution of the microscope is clearly demonstrated. This high spatial resolution is consistent with the estimated 70 nm wavelength of surface plasmons at 502 nm.

Theoretical resolution of such microscope may reach the scale of a few nanometers, since only the Landau damping at plasmon wave vectors of the order of the Fermi momentum seams to be capable of limiting the smallest possible plasmon wavelength. However, increasing resolution may put additional extremely stringent requirements on the quality of the edge of the dielectric microdroplet/mirror used in the microscope and on the surface roughness of the metal substrate. In order to avoid image brightness loss due to plasmon scattering, the edge of the dielectric mirror should be smooth on a scale that is much smaller than the wavelength of the plasmons used. Surface tension of a viscous liquid mitigates this problem to some degree. However, enhancement of the optical resolution down to 10 nm scale may require novel technical solutions. In addition, much work remains to be done on better ways of reconstruction of the original shape of the source from somewhat distorted plasmon-produced images.

Nevertheless, surface plasmon microscope has the potential to become an invaluable tool in medical and biological imaging, where far-field optical imaging of individual viruses and DNA molecules may become a reality. It allows very simple, fast, robust and straightforward image acquisition. Water droplets on a metal surface could be used as elements of 2D optics in measurements where aqueous environment is essential for biological studies (however, the use of water droplets may present some difficulties since change of dielectric media would require different matching conditions with the substrate, and water might not form equally parabolic and stable droplets as glycerin). I should also point out that if used in reverse, surface plasmon immersion microscope may be used in nanometer-scale optical lithography. Both these developments would potentially revolutionize their respective fields.

**Acknowledgement**

The author gratefully acknowledge fruitful collaboration with C.C. Davis, A.V. Zayats and J.Elliott, and support by the NSF grants ECS-0210438 and ECS-0304046.

Figure Captions

Fig.1 (a) Surface Plasmon Immersion Microscope: Surface plasmons are excited by laser light and propagate inside a parabolic-shaped droplet. Placing a sample near the focus of a parabola produces a magnified image in the metal plane, which is viewed from the top by a regular microscope. Used in reverse, this configuration may be used in subwavelength optical lithography. (b) Sketch of the Ar-ion laser lines positions with respect to the dispersion curve of plasmons on the gold-glycerine interface. At 502 nm glycerine has a very large effective refractive index for surface plasmons. Also shown are the approximate locations of other guided optical modes inside the thin layer of glycerine.

Fig.2 The dispersion laws of surface plasmons and normal photons propagating inside the dielectric at small angles along the metal-dielectric interface. *K*-axis represents the quasi-momentum of the respective electromagnetic mode. The intersections between the modes are shown by the dots. Plasmon propagation length around these intersections is expected to increase.

Fig.3 (a) Exponentially decaying surface plasmon beam emitted from an artificial pinhole in the 50 nm thick gold film immersed in a thin glycerin droplet stained with the bodipy die. The cross section of the beam is shown in (b). The plasmon propagation length appears to be of the order of 3 micrometers at 502 nm. Image (c) and its cross-section (d) show the effect of mode coupling due to the slowly varying shape of the glycerin droplet: quickly decaying surface plasmon beams emitted by two pinholes give rise to weaker guided mode beams, which exhibit much longer propagation length. Sketch in (e) illustrates how the mode coupling effect may conserve angular resolution.

Fig.4 The glycerin droplets were formed in desired locations by bringing a small probe (a) wetted in glycerin into close proximity to a sample. The probe was prepared from a tapered optical fiber, which has an epoxy microdroplet near its apex. Bringing the probe to a surface region covered with glycerin led to a glycerin microdroplet formation (b) under the probe in locations indicated by the arrows.

Fig.5 2D images of a 30x30 $\mu m^2$ rectangular nanohole array with 500 nm hole spacing, which are formed in various droplets. The arrow in (a) indicates the droplet edge visible due to plasmon scattering. Approximate theoretical reconstructions of the images via ray tracing are shown to the right of each experimental image. Individual nanoholes of the arrays are shown as individual dots in the theoretical images. 10x microscope objective was used in obtaining images (a) and (c), while 50x objective was used in (e). Comparison of (e) and (f) indicates that individual nanoholes are resolved in the image (e) obtained at 502 nm.

Fig.6 Resolution test of the microscope. The array of triplet nanoholes (c) is imaged using a glycerine droplet shown in (a) using a 10x microscope objective. The image of the triplet array obtained using a 100x objective at 515 nm is shown in (b). The least-distorted part of the image (b) is shown at higher digital zooms of the CCD camera mounted onto the microscope in (e) and (f). Comparison of the image (b) with the theoretically calculated image (d) clearly prove the resolving of the triplet structure.

Fig.7 (a) Image magnification measured in the surface plasmon image of the triplet nanohole array along the line shown in the inset, which is parallel to the optical axis of the droplet. The dots in the graph show the distance between the neighbouring triplets in the image as a function of the triplet position measured along the optical axis. (b) The cross section through the line of double holes in the image of the triplet array.

Fig.8 Images of the gaps in the 30x30 $\mu m^2$ periodic nanohole array. One of the gaps is indicated by an arrow in the electron microscope image of the structure (a). Similar but wider gaps are seen in the plasmon microscope image (b), it's theoretical ray-optics reconstruction (c), and the cross section of the plasmon image (d) obtained along the line shown in (b).

Fig.9 Evaluation of the microscope resolution at 502 nm. The triplets visible at largest magnification are indicated by the arrows in the measured (a) and calculated (b) images. The same triplets are shown at a higher zoom in the experimental image (c). The cross section through two individual nanoholes in the triplet along the line shown in (c) is presented in (d).

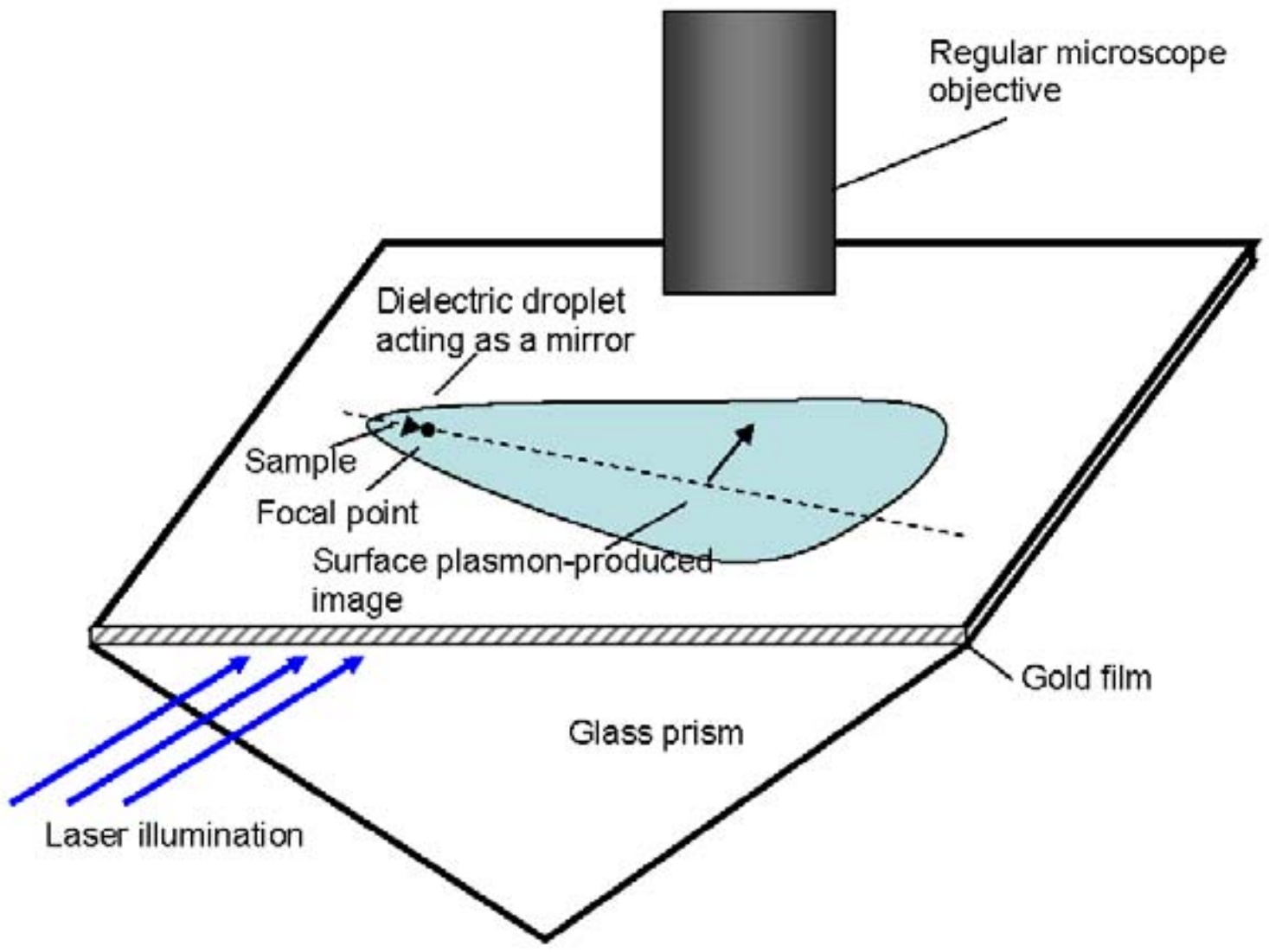

(a)

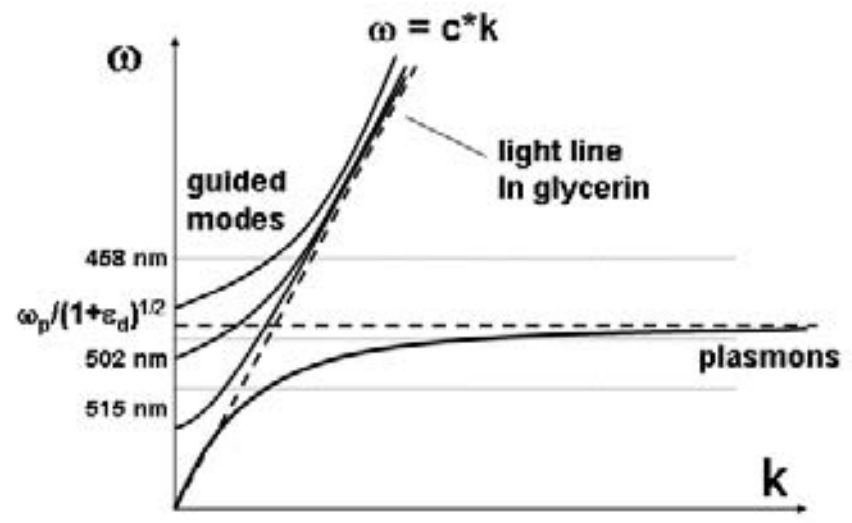

(b)

Fig.1

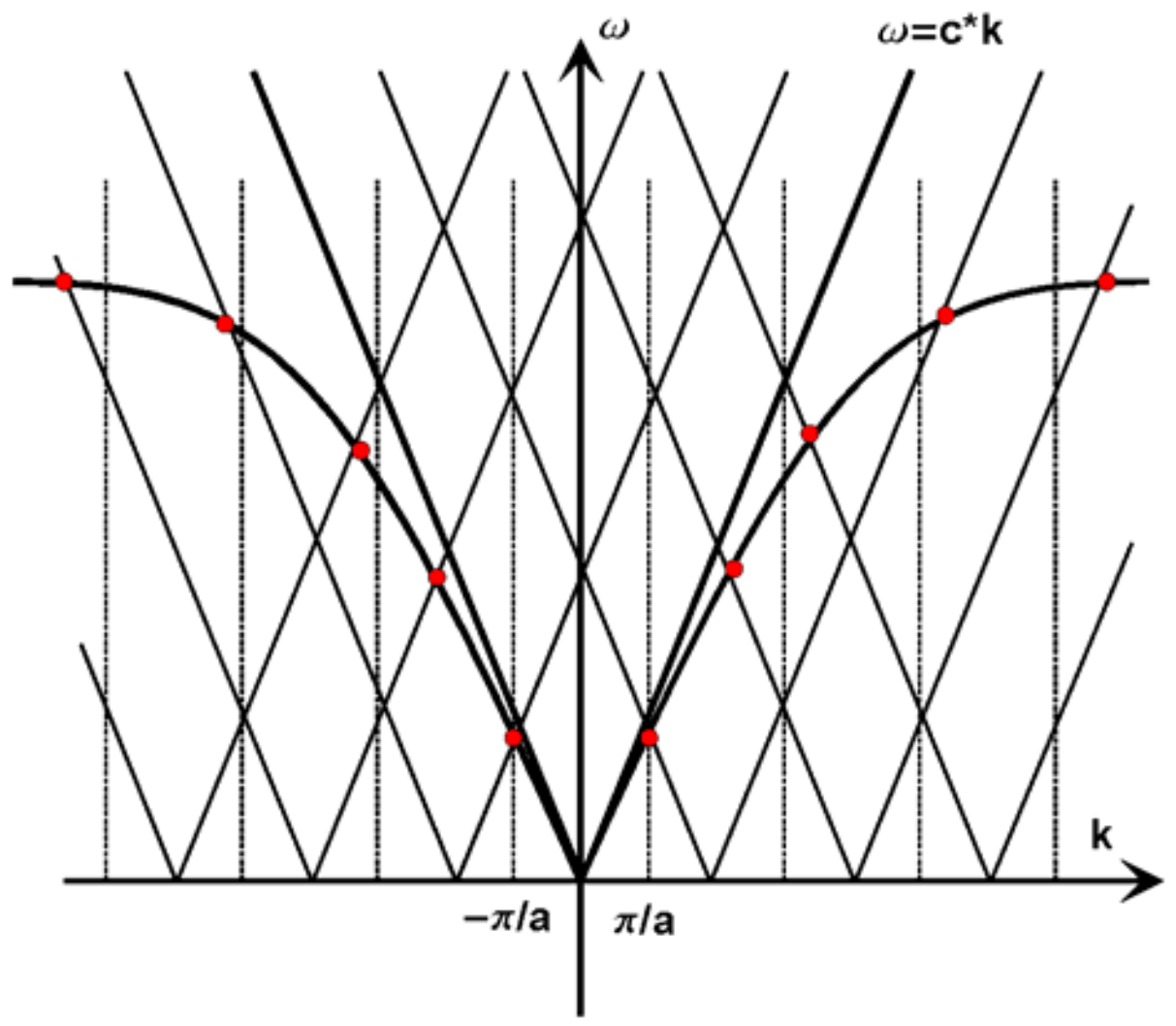

Fig.2

27

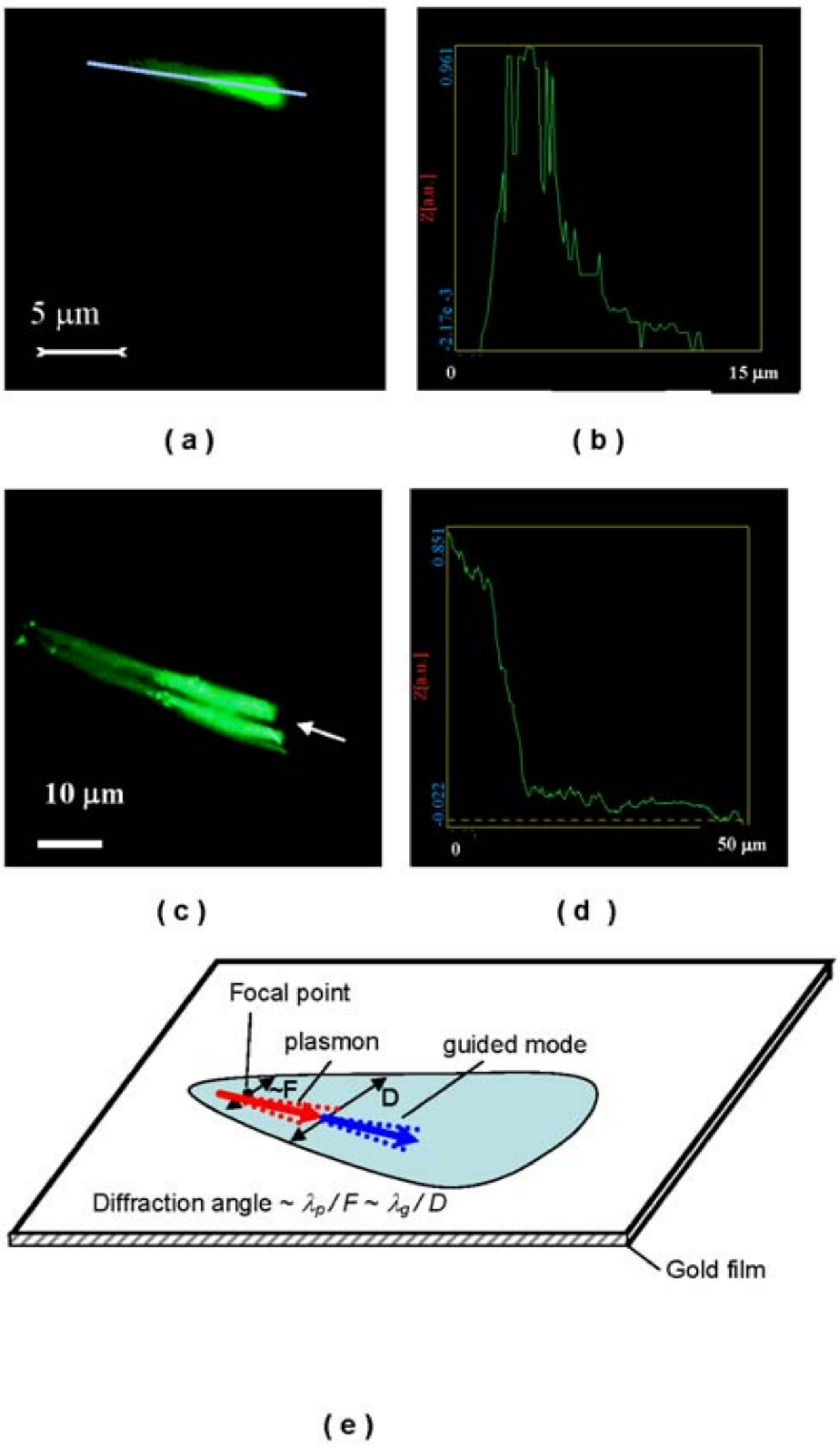
5 μm
0.961
Z[a.u.]
-2.17e-3
0
15 μm
(a)
(b)
10 μm
0.851
Z[a.u.]
-0.022
0
50 μm
(c)
(d)
Focal point
plasmon
guided mode
F
D
Diffraction angle ~ $\lambda_p / F$ ~ $\lambda_g / D$
Gold film
(e)

Fig.3

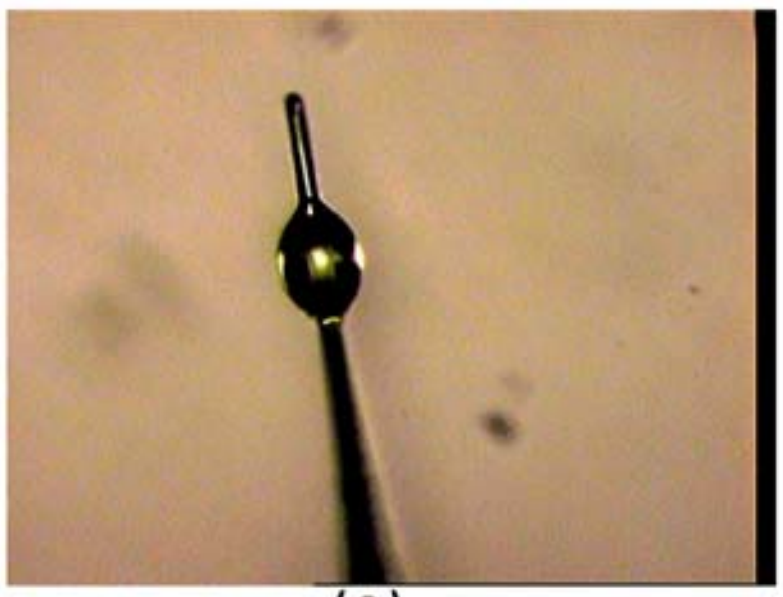

( a )

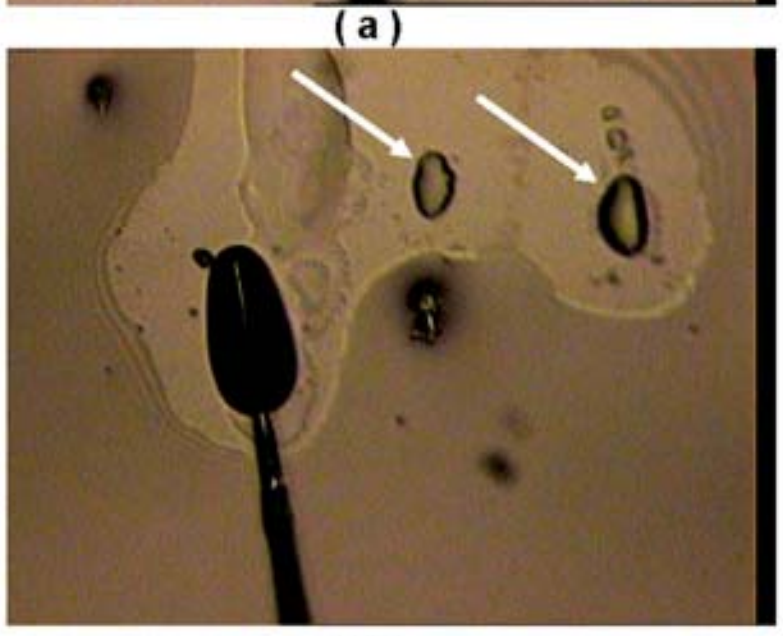

( b )

Fig.4

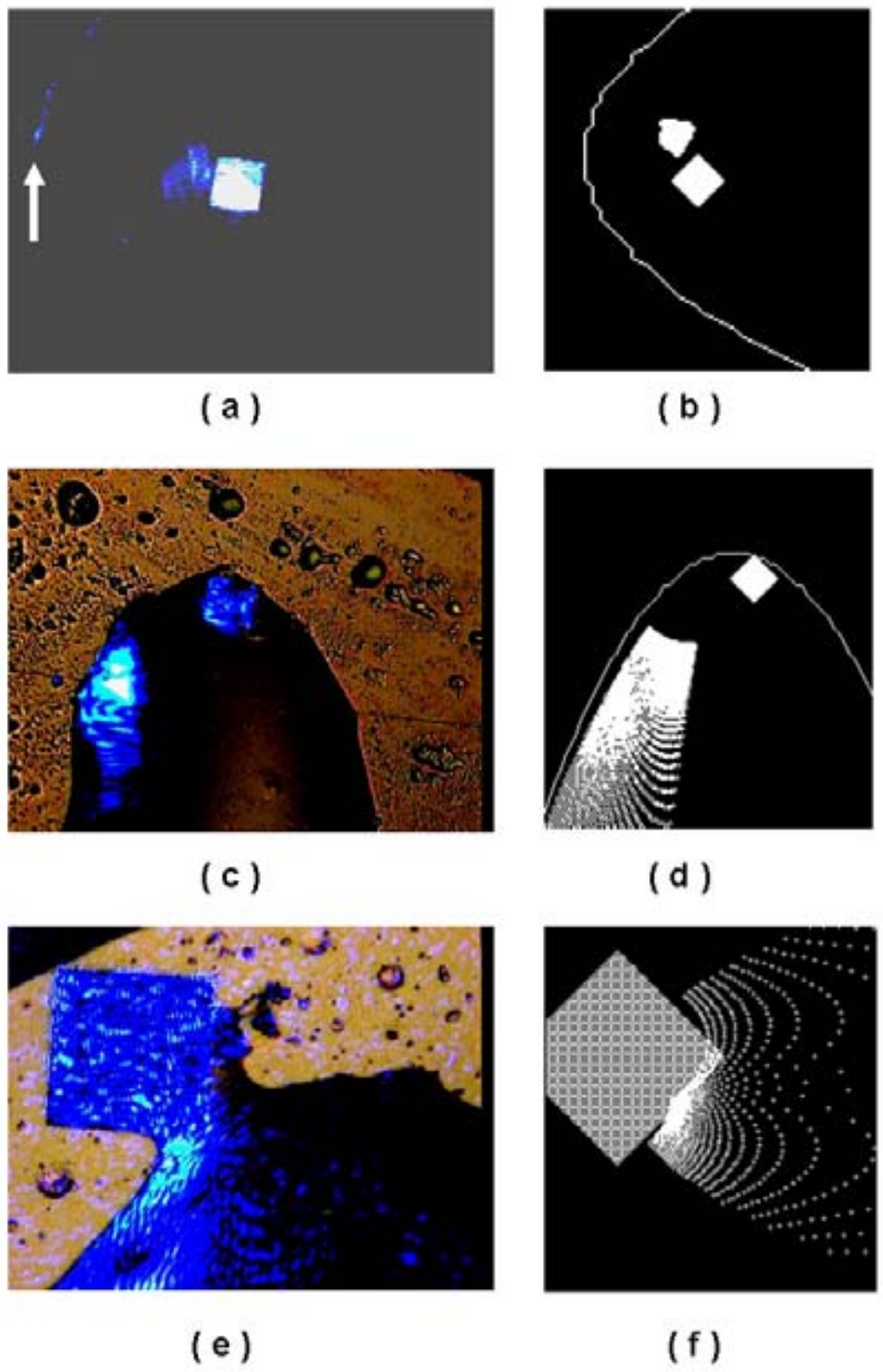

Fig.5

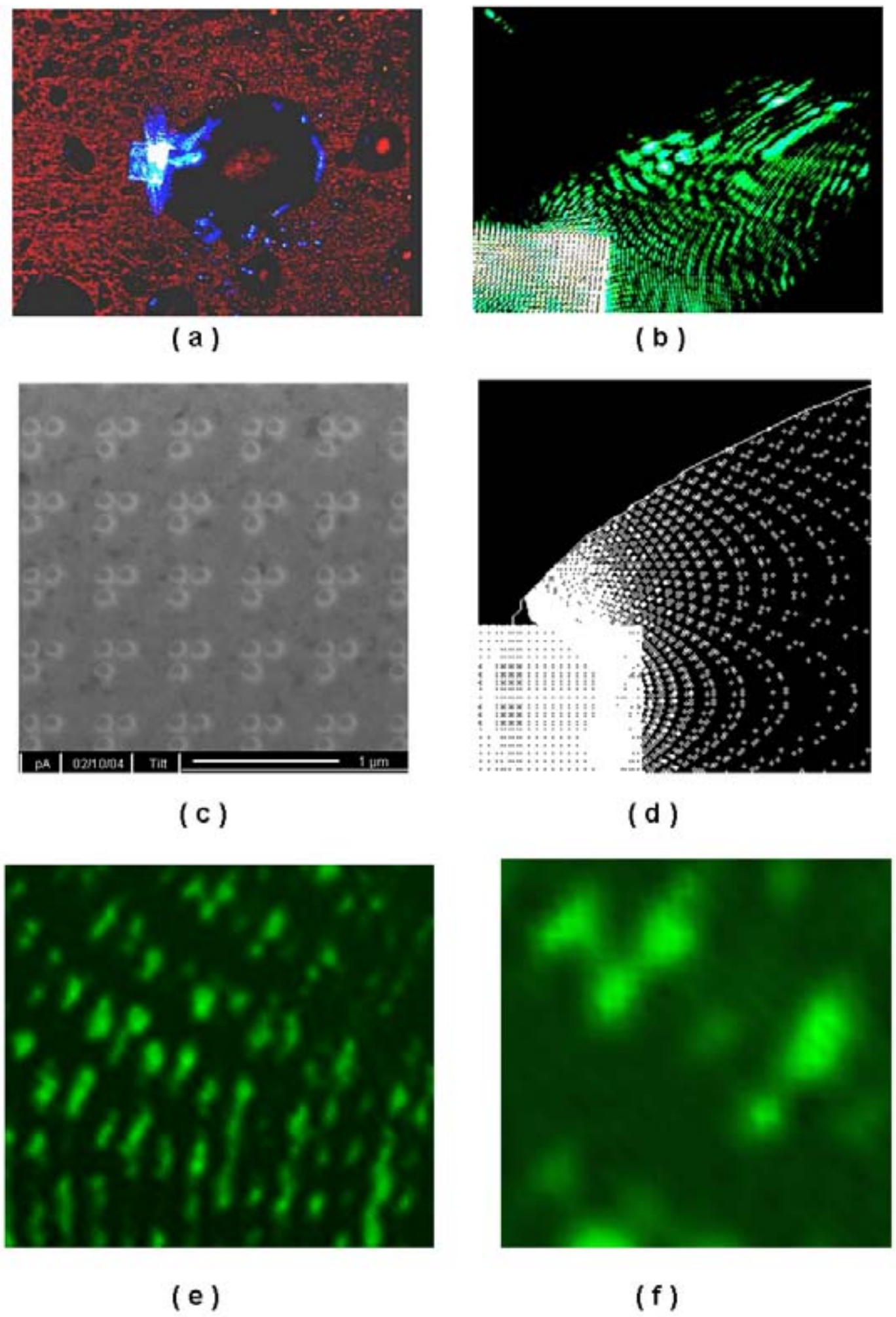

Fig. 6

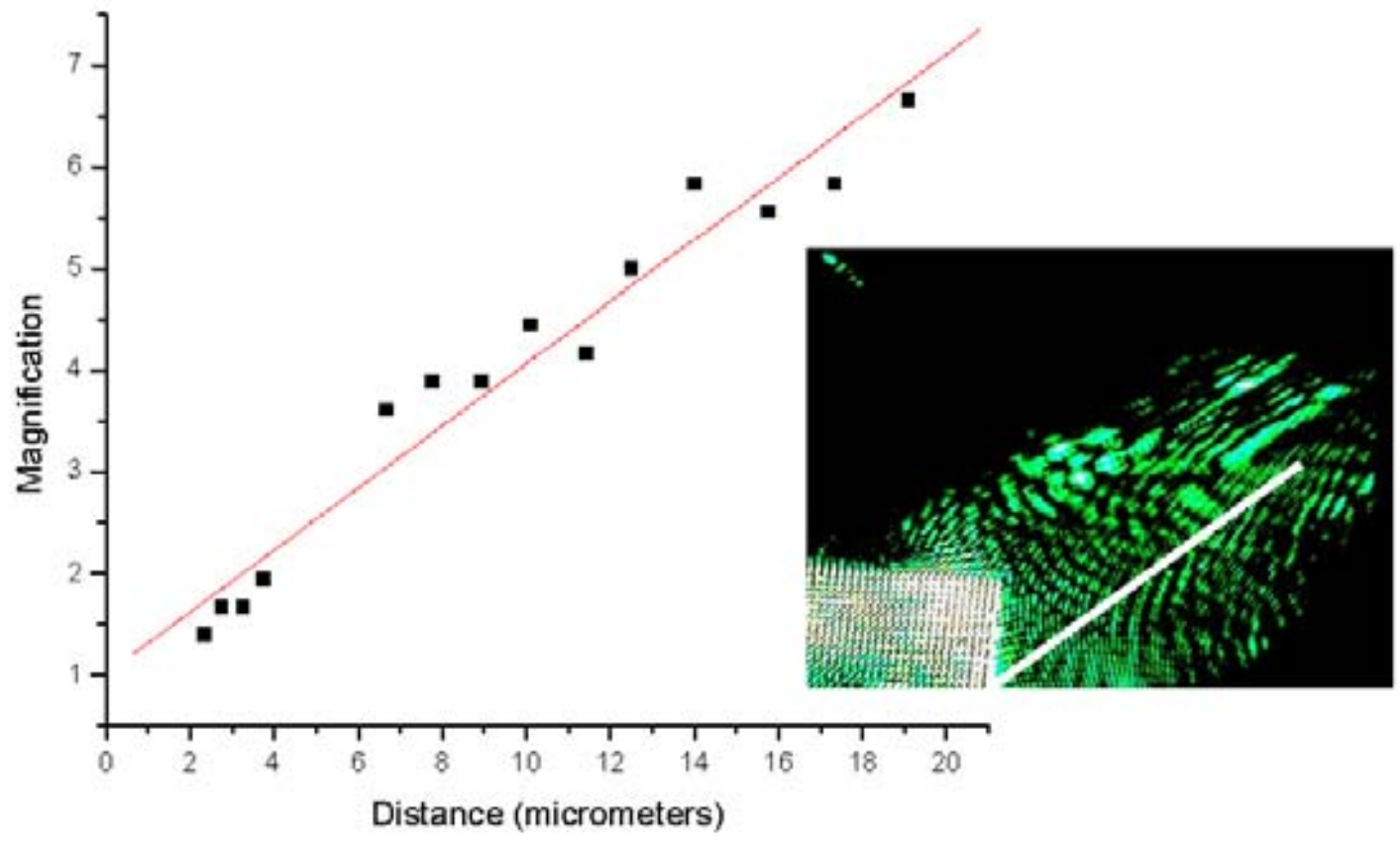

( a )

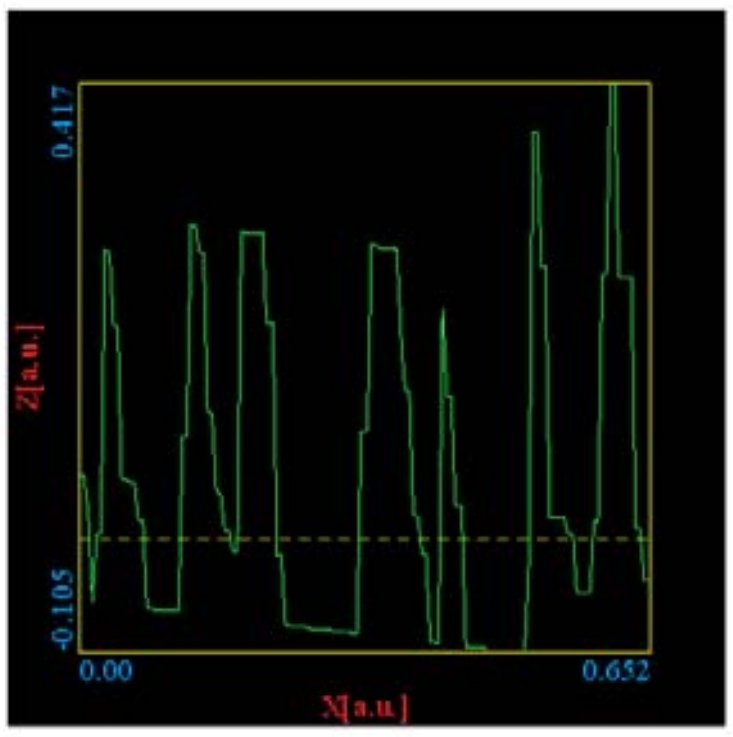

( b )

Fig.7

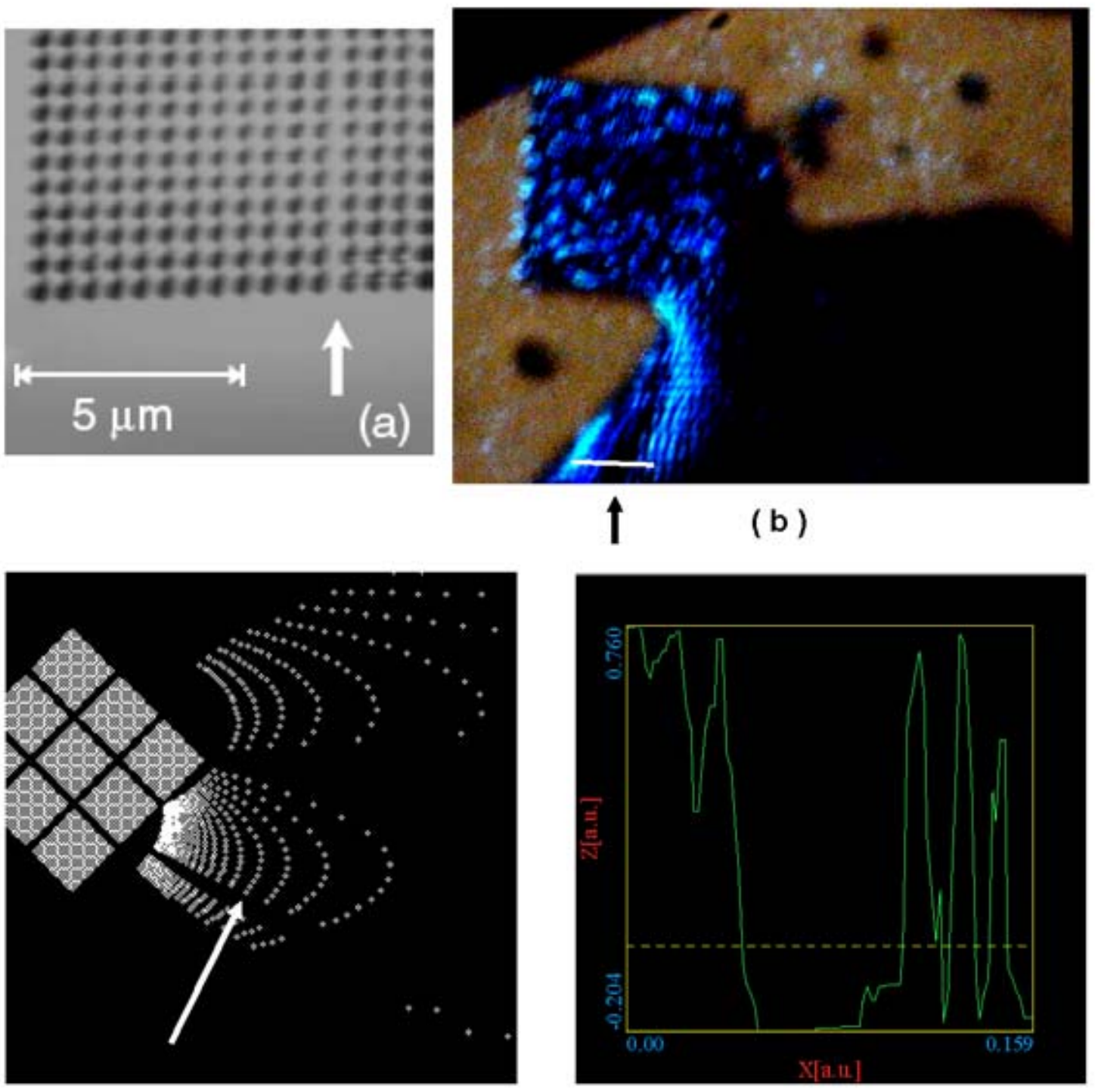

(c) (d)

Fig. 8

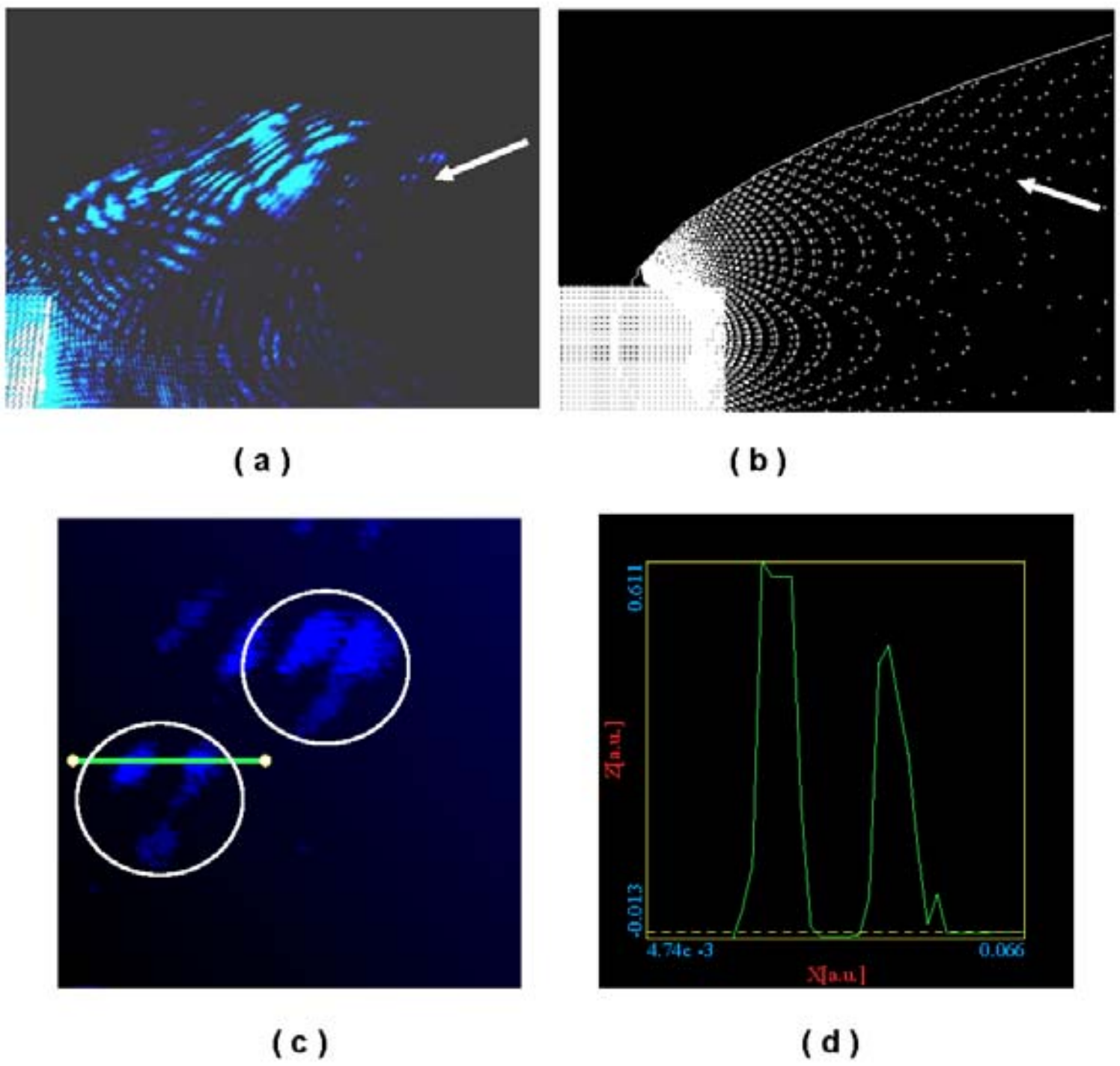

Fig.9